\documentclass[pra,aps,amsmath,amssymb,twocolumn]{revtex4-1}

\usepackage{mathptmx}       
\usepackage{helvet}         
\usepackage{courier}        
\usepackage{type1cm}        
\usepackage{makeidx}         
\usepackage{graphicx}        
\usepackage[bottom]{footmisc}
\usepackage{textcomp}        

\newcommand{\trac}{{\rm Tr}}

\newcommand{\diag}{{\rm diag}}

\newcommand{\be}{\begin{equation}}
\newcommand{\ee}{\end{equation}}
\newcommand{\bea}{\begin{eqnarray}}
\newcommand{\eea}{\end{eqnarray}}
\newcommand{\lt}{\left(}
\newcommand{\rt}{\right)}
\newcommand{\lgr}{\left\{}
\newcommand{\rgr}{\right\}}

\newcommand{\onehalf}{{\mbox{\textonehalf\,}}}
\newcommand{\quot}[1]{{``#1''}}

\begin{document}

\title{ Steepest-Entropy-Ascent and Maximal-Entropy-Production Dynamical Models of Irreversible Relaxation to Stable Equilibrium from Any Non-Equilibrium State.\\ Unified Treatment for Six Non-Equilibrium Frameworks }

\bibliographystyle{unsrt}



\author{Gian Paolo Beretta}

\affiliation{
Universit\`a di Brescia, via Branze 38, 25123
Brescia, Italy, Email: beretta@ing.unibs.it}

\begin{abstract}

By suitable reformulations, we review the mathematical frameworks of six different approaches to the description of non-equilibrium dynamics with the purpose to set up a unified formulation of the  Maximum Entropy Production (MEP) principle  valid in all these contexts. In this way, we extend to such  frameworks  the concept of Steepest Entropy Ascent dynamics introduced  by the present author in previous work on quantum thermodynamics. Actually, the present formulation constitutes a generalization also in the quantum thermodynamics framework. The  analysis emphasizes that in the SEA-inspired implementation of the MEP principle, a key role is played by the geometrical metric with respect to which to measure the length of a trajectory in state space. The metric tensor turns out to be directly related to the inverse of the Onsager's generalized conductivity tensor.  We  conclude that in most of the existing theories of non-equilibrium the time evolution of the state representative can be seen to actually follow in state space the path of SEA with respect to a suitable metric connected with the generalized conductivities.
The resulting unified family of  SAE/MEP dynamical models are all intrinsically  consistent with the second law of thermodynamics. The nonnegativity of the  entropy production  is a general and readily proved feature of SEA dynamics. In  several of the different approaches  to non-equilibrium description we consider here, the SEA concept has not been investigated before. Therefore, it is hoped that the present unifying approach may prove useful in providing a fresh basis for effective, thermodynamically consistent, numerical models and theoretical treatments of irreversible conservative relaxation towards equilibrium from far non-equilibrium states. The six mathematical frameworks are: A) Classical Statistical Mechanics; B)  Small-Scale and Rarefied Gases Dynamics (i.e., kinetic models for the Boltzmann equation); C) Statistical or Information Theoretic Models of Relaxation; D) Rational Extended Thermodynamics, Macroscopic Non-Equilibrium Thermodynamics, and Chemical Kinetics; E) Mesoscopic Irreversible Thermodynamics; F) Quantum Statistical Mechanics, Quantum Thermodynamics, Mesoscopic Non-Equilibrium Quantum Thermodynamics, and Intrinsic Quantum Thermodynamics.
\end{abstract}

\maketitle

\section*{INTRODUCTION}

The problem of understanding entropy and irreversibility has been
tackled by a large number of preeminent scientists during the past
century. Schools of thought have formed and flourished around
different perspectives of the problem. Several modeling approaches have been developed in various frameworks to deal with the many facets of non-equilibrium.

In this paper, we show how to construct Steepest Entropy Ascent (SEA) and Maximum Entropy Production  (MEP) models of non-equilibrium dynamics by adopting a unified mathematical formulation that allows us to do it at once in several different well-known frameworks of non-equilibrium description.

To avoid doing inevitable injustices to the many pioneers of all these approaches and to the many and growing fields of their application, here we skip a generic introduction and given no references nor a review of previous work. Rather, we dig immediately into the mathematical reformulations of the different frameworks in such a way that then the construction of the proposed SEA dynamics becomes formally a single geometrical problem that can be treated at once.

 Our reformulations here not only  allow  a unified treatment of the MEP principle (for a recent review see \cite{Martyushev}) in the various frameworks, but also extends to all frameworks an observation that we have been developing in the quantum thermodynamics framework for the last three decades \cite{Frontiers,LectureNotes,Gheorghiu,ROMP}. In doing so, we introduce an important generalization also in the quantum thermodynamics framework.

 The observation is that we cannot simply  maximize the entropy production  subject to a set of conservation constraints or boundary conditions, but in order to identify a SEA path in state space we must  equip it with a metric with respect to which to compute the distance traveled in state space during the time evolution.

 The generalization is as follows.
  In our previous work, we adopted the proper uniform metric for probability distributions, namely, the Fisher-Rao metric, because in quantum thermodynamics the state representative, the density operator, is essentially a generalized probability distribution. In other frameworks, however, the state representative not always is a probability distribution. Moreover, the present application to the framework of Mesoscopic Non-Equilibrium Thermodynamics \cite{Mazur,BedeauxMazur} shows that standard results such as the Fokker-Planck equation and Onsager theory emerge as straightforward results of SEA/MEP dynamics with respect to a metric characterized by a generalized metric tensor that is directly related to the inverse of the generalized conductivity tensor. Since the generalized conductivities represent, at least in the near-equilibrium regime, the strength of the system's reaction when pulled out of equilibrium, it appear that their inverse, i.e., the generalized resistivity tensor, represents the metric with respect to which the time evolution, at least in the near equilibrium, is locally SEA/MEP.

  But the local SEA/MEP construction does much more, because it offers a strongly thermodynamically consistent way to extend the well-known near-equilibrium theories to the treatment of non-equilibrium states.

An investigation of the interrelations between the SEA and MEP concepts and Ziegler's \cite{Ziegler} and Edelen's \cite{Edelen} formulations for the study of  highly non-equilibrium dynamics in the nonlinear domain is under way and will be communicated elsewhere.

  The unified formulation of the local SAE/MEP variational problem is as follows and it is not restricted to near equilibrium: \textit{the
  time evolution and transport equations advance the local state representative in the direction of maximal  entropy production per unit of distance  traveled in state space compatible with the conservation constraints.}
The measure of distance traveled in state space requires the choice of a metric defined over the state space. The standard near-equilibrium results obtain when the local  metric tensor is proportional to the inverse of the local matrix of generalized conductivities.

In the next six sections we introduce slightly nonstandard notations in several non-equilibrium contexts with the purpose to formulating, in the seventh section,  a unified construction and implementation of  the SAE/MEP concept.

\section*{FRAMEWORK A: CLASSICAL STATISTICAL MECHANICS}

Let $ \Omega $ be the classical position-momentum $q$--$p$ phase space, and $ \mathcal{L} $ the set of
real, square-integrable functions $ A , B , \ldots $ on $ \Omega
$, equipped with the inner product $ ( \cdot | \cdot ) $ defined
by \be (A | B) = \trac (A B) = {\textstyle \int_{\Omega}}\; A B \,\,
d \Omega \label{1a} \ee \noindent where $ \trac (\cdot) $ in this
framework denotes $ \int_{\Omega} \cdot\,d \Omega $, with $d \Omega =d\textbf{q}\,d\textbf{p}$.


In Classical Statistical Mechanics, the index
of statistics from a generally heterogeneous ensemble of identical
systems (with associated phase space $ \Omega $) distributed over
a range of possible classical mechanical states is represented by a  nonnegative (Gibbs)
density-of-phase distribution function $ f_{\rm G}=f_{\rm G}(\textbf{q},\textbf{p},t) $ in $ \mathcal{L} $.

Borrowing from the formalism we originally developed for the quantum framework \cite{Frontiers,LectureNotes} (later introduced also in \cite{Gheorghiu,Reznik}), in order to easily impose the constraint of preservation of the nonnegativity of $f_{\rm G}$ during its time evolution, we adopt as state representative not $f_{\rm G}$ itself but its square root, that we assume is a function in $ \mathcal{L} $ that denote by $ \gamma=\gamma(\textbf{q},\textbf{p},t) $. Normalization is not imposed at this stage but later as one of the constraints.  Therefore, we clearly have
\be f_{\rm G} = \gamma^2 \ ,\qquad \frac{\partial f_{\rm G}}{\partial t}=2\gamma\frac{\partial\gamma}{\partial t}\ee \be  \frac{\partial f_{\rm G}}{\partial \textbf{q}}=2\gamma\frac{\partial\gamma}{\partial \textbf{q}}\ ,\quad \frac{\partial f_{\rm G}}{\partial \textbf{q}}=2\gamma\frac{\partial\gamma}{\partial \textbf{q}}\ ,\quad  \{H,f_{\rm G}\} = 2\gamma \{H,\gamma\} \ee where
$\{\cdot,\cdot\}$ denotes the Poisson bracket.

Among the phase-space functions that represent physical observables we focus on the conserved ones that we denote synthetically by the set \be\lgr C_i \rgr =
\lgr H , M_x , M_y , M_z, N_1 , \ldots , N_r, I \rgr \ee
where $ H $ is the classical Hamiltonian function, $M_j$ the momentum function for the $j$-th component, $ N_i $ the
number-of-particle function for particles of type $ i $, and $I=1$ is the constant unity function, so that $\trac(\gamma^2 H)$ represents the mean energy, $\trac(\gamma^2 \textbf{M})$ the mean momentum vector,  $\trac(\gamma^2 N_i)$  the mean number of particles of type $ i $, and $\trac(\gamma^2 I)$  the normalization condition on $f_{\rm G}$.

The description of an irreversible diffusion-relaxation process in this framework can be done by assuming a evolution equation for the state $f_{\rm G}$ given by
\be \frac{d\gamma}{dt} = \Pi_{\gamma} \quad {\rm where}\quad \frac{d}{dt} = \frac{\partial}{\partial t} -\{H,\cdot\} \label{3a} \ee
It is easy to verify that for $\Pi_{\gamma}=0$ Eq.\ (\ref{3a}) reduces to Liouville's equation of classical reversible evolution. We do not make this assumption because we are interested in modeling irreversible evolution with energy, momentum, and particle numbers redistribution towards equilibrium, subject to the overall conservation of energy, momentum, number of particles of each kind, and normalization
\be \Pi_{C_i}=\frac{d}{dt}\trac(\gamma^2 C_i)=  (2\gamma C_i |\Pi_{\gamma}) = 0 \label{4a} \ee

The entropy state functional in this context is represented by
\be S ( \gamma ) = - k \trac (f_{\rm G} \ln f_{\rm G}) = (-k\gamma\ln\gamma^2 |\gamma)  \label{5a} \ee
so that the rate of entropy production under a time evolution that preserves the normalization of $f_{\rm G}$ is given by
\be \Pi_{S}=- k \frac{d}{dt}\trac(f_{\rm G} \ln f_{\rm G})=  (-2k\gamma\ln \gamma^2 |\Pi_{\gamma})  \label{6a} \ee

Below, in the section on SAE/MEP dynamics, we  construct an equation of motion for the square-root-of-density-of-phase distribution $\gamma$ such that $\Pi_S$ is maximal subject to the conservation constraints $\Pi_{C_i}=0$ and a suitable additional constraint we  discuss therein.

\section*{FRAMEWORK B: SMALL-SCALE AND RAREFIED GASES DYNAMICS}

Let $ \Omega_c $ be the classical one-particle velocity space, and $ \mathcal{L} $ the set of
real, square-integrable functions $ A , B , \ldots $ on $ \Omega_c
$, equipped with the inner product $ ( \cdot | \cdot ) $ defined
by \be (A | B) = \trac (A B) = {\textstyle \int_{\Omega_c}} A B \,\,
d \Omega_c \label{1b} \ee \noindent where $ \trac (\cdot) $ in this
framework denotes $ \int_{\Omega_c} \cdot\,d \Omega_c $, with $d \Omega_c =dc_x\,dc_y\,dc_z$.
%

In the Kinetic Theory of Rarefied Gases and Small-Scale Hydrodynamics \cite{Nicolas}, the probability to find a particle at position $\textbf{x}$ with velocity between $\textbf{c}$ and $\textbf{c}+d \textbf{c}$ [where of course $\textbf{c}=(c_x,c_y,c_z)$] is given by  $ f(\textbf{x},\textbf{c},t)\, d \Omega_c / {\textstyle \int_{\Omega_c} } f \, d \Omega_c$ where $ f(\textbf{x},\textbf{c},t) $  is the local phase-density distribution which for every position $\textbf{x}$ and time instant $t$ is a function in  $ \mathcal{L} $.

Also in this framework, in order to easily impose the constraint of preservation of the nonnegativity of $f$ during its time evolution, we introduce the local one-particle state representation not by $f$ itself but by its square root, that we assume is a function in $ \mathcal{L} $ that we denote by $ \gamma=\gamma(\textbf{x},\textbf{c},t) $.
Therefore, we  have
\be f = \gamma^2 \ ,\quad \frac{\partial f}{\partial t}=2\gamma\frac{\partial\gamma}{\partial t}, \quad \frac{\partial f}{\partial \textbf{x}}=2\gamma\frac{\partial\gamma}{\partial \textbf{x}}\ ,\quad \frac{\partial f}{\partial \textbf{c}}=2\gamma\frac{\partial\gamma}{\partial \textbf{c}} \ee

Again, among the velocity-space functions that represent physical observables we focus on the conserved ones that we denote synthetically by the set \be\lgr C_i \rgr =
\lgr H=\onehalf m \textbf{c}\cdot\textbf{c} , M_x=m c_x , M_y=m c_y , M_z=m c_z, m \rgr \ee of functions in $ \mathcal{L}_c
$  where $ H$ is the local kinetic energy function, $ M_x$, $ M_y$, $ M_z$ the components of the local momentum function, and $ m$ the particle mass, so that $\trac(\gamma^2 H)$ represents the local kinetic energy density,  $\trac(\gamma^2 M_i)$  the $i$-th component of the local momentum density, and $\trac(\gamma^2 m)$  the local mass density.

The time evolution of the distribution function $f$ is given by the Boltzmann equation or some equivalent simplified kinetic model equation, which in terms of the square-root distribution may be written in the form
\be \frac{D\gamma}{Dt} = \Pi_{\gamma} \quad {\rm where}\quad \frac{D}{Dt} = \frac{\partial}{\partial t} +\textbf{c}\cdot\frac{\partial}{\partial \textbf{x}} +\textbf{a}\cdot\frac{\partial}{\partial \textbf{c}} \label{3b} \ee
and $\textbf{a}$ denotes the particle acceleration due to external body forces.

In order to satisfy the constraints of  energy, momentum, and mass conservation the collision term $\Pi_{\gamma}$ must be such that
\be \Pi_{C_i}=\frac{\partial \trac(f C_i)}{\partial t}+\nabla\cdot\trac(f \,\textbf{c}\,C_i) =  (2\gamma C_i |\Pi_{\gamma}) = 0 \label{4b} \ee

The local entropy density functional  in this context is represented by
\be S ( \textbf{x},t ) = - k \trac (f \ln f) =  (-k\gamma\ln\gamma^2 |\gamma) \label{5b} \ee
so that the rate of entropy production under a time evolution that preserves the normalization of $f$ is given by
\be \Pi_{S}=- k\frac{\partial \trac(f \ln f)}{\partial t}-k\nabla\cdot\trac(f \,\textbf{c}\,\ln f) = (-2k\gamma\ln \gamma^2 |\Pi_{\gamma})   \label{6b} \ee

Below, in the section on SAE/MEP dynamics, we  construct a new family of models for the collision term $\Pi_{\gamma}$ such that $\Pi_S$ is maximal subject to the conservation constraints $\Pi_{C_i}=0$ and a suitable additional constraint we  discuss therein.

The resulting new family of SEA kinetic  models of the collision integral
in the Boltzmann equation is currently under investigation by comparing it with  standard models such as the well-known BGK model as well as with Monte Carlo simulations of the original Boltzmann
equation for hard spheres \cite{ASME-Nicolas}. In addition to the strong thermodynamics consistency even far from stable equilibrium, Ref. \cite{ASME-Nicolas} gives a proof that in the near-equilibrium limit the SEA model reduces to the BGK
model.

\section*{FRAMEWORK C: STATISTICAL OR INFORMATION THEORETIC MODELS OF RELAXATION TO EQUILIBRIUM}

Let $ \mathcal{L} $ be the set of all $ n \times n $ real,
diagonal matrixes $ A = \diag ( a_j )$, $B = \diag ( b_j )$, \dots
( $n \leq \infty$ ), equipped with the inner product $ ( \cdot |
\cdot ) $ defined by \be \lt A | B \rt = \trac (A B) = {\textstyle
\sum_{j = 1}^n} a_j\, b_j \label{1c} \ee \noindent

In Information Theory \cite{Jaynes}, the probability assignment to a set of $ n $ events, $
p_j $ being the probability of occurrence of the $ j $-th event is represented by $ \rho = \diag ( p_j ) $. Again, in order to easily impose the constraint of preservation of the nonnegativity of the probabilities during their time evolution, we adopt the description in terms of the square-root of $\rho$ that we denote by
\be \gamma = \diag ( \sqrt{p_j} ) \label{2c}\ee

Typically we consider a set of conserved features of the process \be\lgr C_i \rgr= \lgr H , N_1 ,
\ldots , N_r, I \rgr    \ee  of diagonal matrixes $ H = \diag ( e_j )$,
$N_1 = \diag ( n_{1 j } )$, \dots, $N_r = \diag ( n_{rj} ) $, $I = \diag ( 1 ) $ in $
\mathcal{L} $ representing  characteristic features of the
events in the set, which for the $ j $-th event take on respectively the values $ e_j $, $n_{1 j} $, \dots , $n_{r j} $. The corresponding expectation values are $\trac (\rho H)={\textstyle \sum_{j = 1}^n} p_j\,e_j $, $\trac (\rho N_1)={\textstyle \sum_{j = 1}^n} p_j\,n_{1 j}$, \dots, $\trac (\rho N_r)={\textstyle \sum_{j = 1}^n} p_j\,n_{r j}$, and $\trac (\rho I)={\textstyle \sum_{j = 1}^n} p_j = 1$ thus providing the normalization condition on $\rho$.

The time evolution of the square-root probability distribution  $\gamma$ is the solution of the rate equation
\be \frac{d\gamma}{dt} = \Pi_{\gamma} \label{3c} \ee
where in order to satisfy the constraints of conservation of the expectation values $\trac (\rho C_i) $ the  term $\Pi_{\gamma}$ must be such that
\be \Pi_{C_i}=\frac{d}{dt}\trac(\rho C_i)=  (2\gamma C_i |\Pi_{\gamma}) = 0 \label{4c} \ee

The entropy functional  in this context is represented by
\be S ( \gamma ) = - k \trac (\rho \ln \rho) =  (-k\gamma\ln\gamma^2 |\gamma) \label{5c} \ee
so that the rate of entropy production under a time evolution that preserves the normalization of $\rho$ is given by
\be \Pi_{S}=- k \frac{d}{dt}\trac(\rho \ln \rho)=  (-2k\gamma\ln \gamma^2 |\Pi_{\gamma})  \label{6c} \ee

Below, in the section on SAE/MEP dynamics, we  construct a model for the rate term $\Pi_{\gamma}$ such that $\Pi_S$ is maximal subject to the conservation constraints $\Pi_{C_i}=0$ and a suitable additional constraint we  discuss therein.

An attempt along the same lines has been presented in \cite{Lemanska}.

\section*{FRAMEWORK D: RATIONAL EXTENDED THERMODYNAMICS, MACROSCOPIC NON-EQUILIBRIUM THERMODYNAMICS, AND CHEMICAL KINETICS}

Let $ \mathcal{L} $ be the set of all $ n \times n $ real,
diagonal matrixes $ A = \diag ( a_j )$, $B = \diag ( b_j )$, \dots
( $n \leq \infty$ ), equipped with the inner product $ ( \cdot |
\cdot ) $ defined by \be \lt A | B \rt = \trac (A B) = {\textstyle
\sum_{j = 1}^n} a_j\, b_j \label{1d} \ee \noindent

In Rational Extended Thermodynamics \cite{Ruggeri}, the local state at position $\textbf{x}$ and time $t$ of the continuum under study is represented by an element $\gamma$ in $ \mathcal{L} $, i.e.,
\be \gamma(\textbf{x},t)= \diag[\alpha(\textbf{x},t)] \ee

 Thus, $ \gamma(\textbf{x},t) $ represents the  set of fields which represent the instantaneous spatial distributions within the continuum of the local densities that define all its other local properties. In particular, for the  conserved properties energy, momentum, and mass it is assumed  that their local densities and their local fluxes  are all given by particular functions of $\gamma$ that we denote synthetically by \be\lgr C_i (\gamma)  \rgr=
\lgr E(\gamma) , M_x(\gamma) , M_y(\gamma) , M_z(\gamma), m(\gamma) \rgr \ee
\be\lgr \textbf{J}_{C_i}(\gamma)  \rgr =
\lgr  \textbf{J}_E(\gamma) , \textbf{J}_{M_x}(\gamma) , \textbf{J}_{M_y}(\gamma) , \textbf{J}_{M_z}(\gamma), \textbf{J}_{m}(\gamma) \rgr \ee
so that the energy, momentum, and mass balance equations take the form
\be \frac{DC_i}{Dt}=\frac{\partial C_i}{\partial t} +\nabla\cdot\textbf{J}_{C_i} = \Pi_{C_i}= 0  \label{3d} \ee
Moreover, also for the local entropy density and the local entropy flux it  is assumed  that they are given by particular functions of $\gamma$ that we denote respectively by
\be S(\gamma)\qquad {\rm and}\qquad \textbf{J}_{S}(\gamma)\ee
so that the entropy balance equation takes the form
\be \frac{DS}{Dt}=\frac{\partial S}{\partial t} +\nabla\cdot\textbf{J}_S = \Pi_{S}  \label{4d} \ee
where $\Pi_{S}$ is the local production density.

In general the balance equation for each of the underlying field properties is
\be \frac{D\alpha_j}{Dt}=\frac{\partial \alpha_j}{\partial t}+\nabla\cdot\textbf{J}_{\alpha_j}=\Pi_{\alpha_j} \label{6d} \ee
where $\textbf{J}_{\alpha_j}$ and $\Pi_{\alpha_j}$  are the corresponding flux and production density, respectively. Equivalently, this set of balance equations may be written synthetically  as
\be \frac{D\gamma}{Dt}=\frac{\partial \gamma}{\partial t}+\nabla\cdot\textbf{J}_\gamma=\Pi_{\gamma} \label{5d} \ee
where $\textbf{J}_\gamma=\diag [  \textbf{J}_{\alpha_j} ]$  and $\Pi_\gamma=\diag [ \Pi_{\alpha_j} ]$.

It is then further assumed that there exist  functions $\Phi_{\alpha_j}(\gamma)$ (Liu's Lagrange multipliers \cite{Liu}) that we denote here in matrix form by
\be\Phi= \diag( \Phi_{\alpha_j} ) \ee
such that the local entropy production density can be written as
\be \Pi_{S} = \sum_{j=1}^n \Phi_{\alpha_j} \Pi_{\alpha_j}=(\Phi | \Pi_{\gamma})\ee
and must be nonnegative everywhere.

For our development in this paper we shall additionally assume that there also exist  functions $\Psi_{i\,{\alpha_j}}(\gamma)$  that we denote  in vector form by
\be\Psi_i= \diag ( \Psi_{i\,{\alpha_j}})  \ee
such that the production density of each conserved property $C_i$ can be written as
\be \Pi_{C_i} = \sum_{j=1}^n \Psi_{i\,{\alpha_j}} \Pi_{\alpha_j}=(\Psi_i | \Pi_{\gamma})\ee

Typically, but not necessarily, the first five underlying fields $\alpha_j(\textbf{x},t)$ for $j=1,\dots,5$ are conveniently chosen to coincide with the energy, momentum, and mass densities, so that Eqs.\ (\ref{6d}) for  $j=1,\dots,5$ coincide with Eqs.\ (\ref{3d}) because $\Pi_{\alpha_j}=0$ for this subset of conserved fields.

The above framework reduces to the traditional Onsager theory of macroscopic Non-Equilibrium Thermodynamics (NET) \cite{Mazur} if the $\alpha_j$'s are  taken to represent the local deviations  of the underlying fields from their equilibrium values. In this context, the usual notation calls  the functions $X_{\alpha_j}=-\Phi_{\alpha_j}$ the \quot{thermodynamic forces} and $\Pi_{\alpha_j}$ the \quot{thermodynamic currents}.

The same framework reduces to the standard scheme of Chemical Kinetics (CK) if the $\alpha_j$'s are  taken to represent the local reaction coordinates, $\Pi_{\alpha_j}$ the local rate of advancement of reaction $j$, $\Phi_{\alpha_j}$ its entropic affinity, $C_i$ the local concentration of atomic elements of kind $i$, $\Pi_{C_i}=0$ their local production density.

Below, in the section on SAE/MEP dynamics, we  construct an equation of motion for $\gamma$ such that $\Pi_S$ is maximal subject to the conservation constraints $\Pi_{C_i}=0$ and a suitable additional constraint we  discuss therein.

 \section*{FRAMEWORK E. MESOSCOPIC NON-EQUILIBRIUM THERMODYNAMICS}

Let $ \mathcal{L}$ be the set of all $ n \times n $
diagonal matrixes $ A = \diag ( a_j(\gamma) )$, $B = \diag ( b_j(\gamma) )$, \dots whose entries $a_j(\gamma)$, $b_j(\gamma)$, \dots are real, square-integrable functions of a set of mesoscopic properties usually denoted by $\alpha_1,\dots,\alpha_m$ that here we denote synthetically by defining the matrix
\be \gamma = \diag(\alpha_1,\dots,\alpha_m) \label{1e} \ee
 and denoting its $m$-dimensional range by $\Omega_\gamma$, usually called the  $\pmb{\alpha}$-space.
 Let $ \mathcal{L}$ be
equipped with the inner product $ ( \cdot | \cdot ) $ defined
by \be (A | B) = \sum_{i=1}^n \trac (a_i b_i) = \sum_{i=1}^n  \int_{\Omega_\gamma} a_i(\gamma) b_i(\gamma) \,\,
d \Omega_\gamma \label{2e} \ee \noindent where $ \trac (\cdot) $ in this
framework denotes $ \int_{\Omega_\gamma} \cdot\,d \Omega_\gamma $, with $d \Omega_\gamma =d\alpha_1\cdots d\alpha_m$.

 In Mesoscopic Non-Equilibrium Thermodynamics (MNET)  (see, e.g., \cite{Mazur})  the  $\alpha_j$'s are the set of mesoscopic (coarse grained) local extensive properties assumed to  represent the local non-equilibrium state of the portion of continuum under study. The mesoscopic  description of the local state at position $\textbf{x}$ and time $t$  is in terms of a probability density  on the $\pmb{\alpha}$-space $\Omega_\gamma$, that we denote by $P(\gamma;\textbf{x},t)$. $P(\gamma;\textbf{x},t)\,d \Omega_\gamma$ represents the probability that the values of the underlying fields are between $\gamma$ and $\gamma+d\gamma$.

 It is assumed that the probability density $P$  obeys a continuity equation that we may write as follows
 \be \frac{DP}{Dt}=\frac{\partial P}{\partial t}+\textbf{c}\cdot  \nabla P=-\nabla_\gamma\cdot \Pi_\gamma \label{3e} \ee
 where $\textbf{c}=\textbf{c}(\gamma)$ is the particle velocity expressed in terms of the underlying fields (usually it is convenient to take the first three $\alpha_j$'s to coincide with the velocity components) and  we define for shorthand
 \be \Pi_\gamma =  \diag ( \Pi_{\alpha_j} ) \quad {\rm and}\quad \nabla_\gamma=\diag \left( \frac{\partial}{\partial\alpha_j} \right) \label{4e}\ee
where the $\Pi_{\alpha_j}$'s are interpreted as the components of a streaming flux in $\Omega_\gamma$, i.e., a current in the space of mesoscopic coordinates.

 The conserved fields $C_i(\textbf{x},t)$ have an associated underlying extensive property which can be expressed in terms of the mesoscopic coordinates as $\psi_i(\gamma)$. They obey the balance equation
 \be \frac{DC_i}{Dt}=\frac{\partial C_i}{\partial t} +\nabla\cdot\textbf{J}_{C_i} = \Pi_{C_i}= 0  \label{5e} \ee
 where local density $C_i(\textbf{x},t)$, the local flux $\textbf{J}_{C_i}(\textbf{x},t)$ and the local production density $\Pi_{C_i}(\textbf{x},t)$ are defined as follows
  \bea C_i(\textbf{x},t)  &=& \int_{\Omega_\gamma} \psi_i(\gamma)\,P(\gamma;\textbf{x},t)\,d \Omega_\gamma  \nonumber\\
  \textbf{J}_{C_i}(\textbf{x},t)  &=& \int_{\Omega_\gamma} \psi_i(\gamma)\,\textbf{c}(\gamma)\,P(\gamma;\textbf{x},t)\,d \Omega_\gamma \nonumber\\
  \Pi_{C_i}(\textbf{x},t)  &=& \int_{\Omega_\gamma} \psi_i(\gamma)\,\frac{DP}{Dt}(\gamma;\textbf{x},t)\,d \Omega_\gamma \nonumber\\
 &=&-
  \int_{\Omega_\gamma} \psi_i(\gamma)\,\nabla_\gamma\cdot \Pi_\gamma\,d \Omega_\gamma
 \nonumber\\
 &=&
 \int_{\Omega_\gamma} \Pi_\gamma\cdot \nabla_\gamma\psi_i(\gamma)\, \,d \Omega_\gamma
 \nonumber\\
 &=& (\Psi_i| \Pi_\gamma)
 \label{6e}
  \eea
  where in the next to the last equation we integrated by parts and assumed that currents in $\pmb{\alpha}$-space decay sufficiently fast to zero as the  $\gamma_j$'s $\rightarrow\infty$, and  we defined
  \be \Psi_i= \nabla_\gamma\psi_i(\gamma) \label{6e1}\ee

The entropy balance equation takes the form
\be \frac{DS}{Dt}=\frac{\partial S}{\partial t} +\nabla\cdot\textbf{J}_S = \Pi_{S}  \label{7e} \ee
where the local density $S(\textbf{x},t)$, the local flux $\textbf{J}_{S}(\textbf{x},t)$ and the local production density $\Pi_{S}(\textbf{x},t)$ are defined in terms of the associated extensive property expressed in terms of the mesoscopic coordinates as
\be \phi(\gamma) = - k \ln P(\gamma) \label{8e}\ee
 as follows
\bea S(\textbf{x},t)  &=& \int_{\Omega_\gamma} \phi(\gamma)\,P(\gamma;\textbf{x},t)\,d \Omega_\gamma  \nonumber\\
  \textbf{J}_{S}(\textbf{x},t)  &=& \int_{\Omega_\gamma} \phi(\gamma)\,\textbf{c}(\gamma)\,P(\gamma;\textbf{x},t)\,d \Omega_\gamma \nonumber\\
  \Pi_{S}(\textbf{x},t)  &=& \int_{\Omega_\gamma} \phi(\gamma)\,\frac{DP}{Dt}(\gamma;\textbf{x},t)\,d \Omega_\gamma \nonumber\\
 &=&-
  \int_{\Omega_\gamma} \phi(\gamma)\,\nabla_\gamma\cdot \Pi_\gamma\,d \Omega_\gamma
 \nonumber\\
 &=&
 \int_{\Omega_\gamma} \Pi_\gamma\cdot \nabla_\gamma\phi(\gamma)\, \,d \Omega_\gamma
\nonumber\\
 &=& (\Phi| \Pi_\gamma)
  \label{9e}
  \eea
 where again in the next to the last equation we integrated by parts and we defined
 \be \Phi= \nabla_\gamma\phi(\gamma) \label{10e}\ee

Below, in the section on SAE/MEP dynamics, we  construct an equation of motion for $\gamma$ such that $\Pi_S$ is maximal subject to the conservation constraints $\Pi_{C_i}=0$ and a suitable additional constraint we  discuss therein. The result, when introduced in Eq.\ (\ref{3e}) will yield the Fokker-Planck equation for $P(\gamma;\textbf{x},t)$ which is also related (see, e.g., \cite{GorbanKarlin}) to the GENERIC structure \cite{GENERIC}. The formalism can also be readily extended to the family of Tsallis \cite{Tsallis} entropies in the frameworks of non-extensive thermodynamic models \cite{Gorban}.

\section*{FRAMEWORK F: QUANTUM STATISTICAL MECHANICS, QUANTUM INFORMATION THEORY, QUANTUM THERMODYNAMICS, MESOSCOPIC NON-EQUILIBRIUM QUANTUM THERMODYNAMICS, AND INTRINSIC QUANTUM THERMODYNAMICS}

Let $ \mathcal{H} $ be the Hilbert space (dim $ \mathcal{H} \leq
\infty $) associated with the physical system, and $ \mathcal{L} $ the set of all linear operators $ A
$, $B $, \dots  on $ \mathcal{H} $, equipped with the real inner
product $( \cdot | \cdot )$ defined by \be \lt A | B \rt = \trac
\lt A^{\dag} B + B^{\dag} A \rt/2 \label{1f} \ee \noindent where $
A^{\dag} $ denotes the adjoint of operator $ A $ and $ \trac(
\cdot ) $ the trace functional.

In the quantum frameworks that we consider in this section, the state representative is the density operator $\rho$, i.e.,  a unit-trace, self-adjoint, and nonnegative-definite element of $ \mathcal{L} $.

Instead, also here we will adopt  the state representation in terms of the generalized square root of the density operator, that we developed in this context \cite{Frontiers,LectureNotes,Gheorghiu,ROMP} in order to easily impose the constraints of preservation of both the nonnegativity and the self-adjointness of $\rho$ during its time evolution. Therefore, we assume that the state representative is an element $\gamma$ in $ \mathcal{L} $ from which we can compute the density operator as follows
\be \rho = \gamma\gamma^\dagger \ee
In other words,
 we adopt as state representative not the density operator $\rho$ itself but its generalized square root $\gamma$. Therefore, we clearly have
\be \frac{d\rho}{d t}=\gamma\frac{d\gamma\dagger}{d t}+\frac{d\gamma}{d t}\gamma^\dagger\label{2f}\ee

We  then
consider the set of operators corresponding to the conserved properties, denoted synthetically as \be\lgr C_i \rgr= \lgr H , M_x, M_y, M_z, N_1 , \ldots , N_r, I \rgr \label{3f}\ee Here we assume that these are self-adjoint operators in $ \mathcal{L} $, that each $M_j$ and $ N_i $
commutes with $ H $, i.e., $ H M_j = M_j H $ for  $ j
= x , y , z $ and  $ H N_i = N_i H $ for $ i
= 1 , \ldots , r $, and that $I$ is the identity operator.\footnote{In simplified models, the set $\lgr C_i \rgr$ is often restricted to only $\lgr H , I \rgr$. Operators $M_x$, $M_y$, $M_z$ are the components of the momentum operator.
Operator $ N_i $, for $ i =
1$, \dots , $r $, is the number operator for particles of type $ i
$ in the system. If the system is closed to particle exchange, it has a fixed number $ n_i $ of particles of type $ i $, then $ N_i = n_i I $, i.e., it is a c-number operator, where $ I $ is the identity operator on $ \mathcal{H} $. If the system is open to particle exchange, then the Hilbert space $ \mathcal{H} $ is a Fock space, i.e., $$ \mathcal{H} = \bigoplus_{j_1=0}^\infty \cdots \bigoplus_{j_r=0}^\infty \mathcal{H}_{j_1j_2\dots j_r} \  {\rm    and} \  N_i = \sum_{j_1=0}^\infty \cdots \sum_{j_r=0}^\infty j_i\, I_{j_1j_2\dots j_r} $$ where $ I_{j_1j_2\dots j_r} $ is the projector operator onto the subspace $ \mathcal{H}_{j_1j_2\dots j_r} $ belonging to the composition with $j_1$ particles of type 1, $j_2$ particles of type 2, and so on. }

The semi-empirical description of an irreversible relaxation process is done in this framework by assuming an evolution equation for the state $\gamma$ given by the equations
\bea  \frac{d\gamma}{d t} +\frac{i}{\hbar}\, H\gamma&=& \Pi_\gamma  \label{4f1}\\ \frac{d\gamma^\dagger}{d t} -\frac{i}{\hbar}\,\gamma^\dagger H&=& \Pi_{\gamma^\dagger}  \label{4f2}  \eea
As a result, it is easy to verify that for the density operator the dynamical equation is
\be  \frac{d\rho}{d t} +\frac{i}{\hbar}\, [H,\rho]= \Pi_\gamma\,\gamma^\dagger+\gamma\,\Pi_{\gamma^\dagger} \label{5f}\ee
where $[\cdot,\cdot]$ denotes the commutator. From this
we  see that in order to preserve hermiticity of $\rho$ the dissipative terms  $\Pi_\gamma$ and $\Pi_{\gamma^\dagger}$ must satisfy the conditions
\be \Pi_{\gamma^\dagger}=\Pi^\dagger_\gamma \quad {\rm and}\quad \Pi_{\gamma}=\Pi^\dagger_{\gamma^\dagger}\label{6f}\ee

In order to satisfy the constraints of conservation of the expectation values $\trac (\rho C_i) $, recalling that each $C_i$ commutes with $H$, the  term $\Pi_{\gamma}$ must be such that
\be \Pi_{C_i}=\frac{d}{dt}\trac(\rho C_i)= \trac(C_i \Pi_\gamma\,\gamma^\dagger+\gamma\,\Pi_{\gamma^\dagger}C_i)= (2C_i \gamma|\Pi_{\gamma}) = 0 \label{7f} \ee

The entropy functional  in this context is represented by
\be S ( \gamma ) = - k \trac (\rho \ln \rho) = (-k(\ln \gamma\gamma^\dagger)\, \gamma |\gamma)  \label{8f} \ee
so that the rate of entropy production under a time evolution that preserves the normalization of $\rho$ is given by
\be \Pi_{S}=- k \frac{d}{dt}\trac(\rho \ln \rho)= (-2k(\ln \gamma\gamma^\dagger)\, \gamma |\Pi_{\gamma})  \label{9f} \ee

In Quantum Statistical Mechanics (QSM) and Quantum Information Theory (QIT), $ \rho $ is the von Neumann
statistical or density operator which represents the index of
statistics from a generally heterogeneous ensemble of identical
systems (same Hilbert space $ \mathcal{H} $ and operators $\lgr H
, N_1 , \ldots , N_r \rgr$) distributed over a range of generally
different quantum mechanical  states. If each individual member of
the ensemble is isolated and uncorrelated from the rest of the
universe, its state is described according to Quantum Mechanics by
an idempotent density operator
($\rho^2=\rho=P_{|\psi\rangle}=\frac{|\psi\rangle\langle\psi|}{\langle\psi|\psi\rangle}$),
i.e., a projection operator onto the span of some vector
$|\psi\rangle$ in $ \mathcal{H} $. If the ensemble is
heterogeneous, its individual member systems may be in different
states, $P_{|\psi_1\rangle}$, $P_{|\psi_2\rangle}$, and so on, and the ensemble statistics is captured by the von Neumann statistical operator $\rho=\sum_j w_j P_{|\psi_j\rangle}$. The entropy functional here represents a measure of the informational uncertainty as to  which homogeneous subensemble the next system will be drawn from, i.e., as to which will be the actual pure quantum state among those present in the heterogeneous ensemble.

In this framework, unless the statistical weights $w_j$ change for some extrinsic reason, the quantum evolution of the ensemble is given by Eq.\ (\ref{5f}) with $\Pi_\gamma=0$ so that
 Eq.\ (\ref{5f}) reduces to von Neumann's equation of quantum reversible evolution, corresponding to $\rho(t)=\sum_j w_j P_{|\psi_j(t)\rangle}$ where the underlying pure states $ |\psi_j(t)\rangle$ evolve according to the Schr\"{o}dinger equation $d |\psi_j\rangle/dt=-iH|\psi_j\rangle/\hbar$.

In the framework of QSM and QIT, the SEA  equation of motion we  construct in the next sections for $\rho$ represents a model for the rates of change of  statistical weights $w_j$ in  such a way that $\Pi_S$ is maximal subject to the conservation constraints $\Pi_{C_i}=0$ (and a suitable additional constraint, see below), thus essentially extends to the quantum landscape the same statistical or information theoretic non-equilibrium problem we defined above as Framework C.

In Quantum Thermodynamics (QT), instead, the density operator takes on a more fundamental physical meaning. It is not any longer related to the heterogeneity of the ensemble, and it is not any longer assumed that the individual member systems of the ensemble are in pure states.

 The prevailing interpretation of QT is the so-called open-system model whereby the quantum system under study (each individual system of a homogeneous ensemble) is always viewed as in contact (weak or strong) with a thermal reservoir or 'heat bath', and its not being in a pure state is an indication of its being correlated with the reservoir. The overall system-plus-bath composite is assumed to be in a pure quantum mechanical state $ \mathcal{H}\otimes \mathcal{H}_R $ and reduces to  the density operator $\rho$ on the system's space $ \mathcal{H}$ when we partial trace over the bath's space $ \mathcal{H}_R$.

The semi-empirical description of an irreversible relaxation process is done in this framework by assuming for $\Pi_\rho$ in Eq.\ (\ref{5f}) the  Lindblad-Gorini-Kossakowski-Sudarshan (LGKS) \cite{Lindblad,Kossakowski}
 \be \Pi_\rho =\sum_j \left(V_j\rho V^\dagger_j-\onehalf\{V^\dagger_j V_j,\rho \}\right) \label{10f} \ee
 where $\{\cdot,\cdot\}$ denotes the anticommutator and operators $V_j$ are to be chosen so as to properly model the system-bath interaction. The justification and modeling assumptions that lead to the general form of Eq.\ (\ref{10f}) are well known.

 In the framework of QT the SEA  equation of motion we  construct in the next sections for $\rho$ represents an alternative  model for $\Pi_\rho$ (or for a term additional to the LGKS term) such that $\Pi_S$ is maximal subject to the conservation constraints $\Pi_{C_i}=0$ (and a suitable additional constraint, see below). In some cases this could be  simpler than the LGKS model and it has the advantage of a strong built-in thermodynamics consistency.

Mesoscopic Non-Equilibrium Quantum Thermodynamics (MNEQT) \cite{BedeauxMazur} starts from the formalism of QSM but attempts to extend the Onsager NET theory and  MNET to the quantum realm. We will show elsewhere that the present SEA formulation reduces to MNEQT in the near-equilibrium limit, and can therefore be viewed as the natural extension of MNEQT. The essential elements of this proof have actually already been given \cite{Gheorghiu}, but only for  the particular case corresponding to Eq. (\ref{4g}) below (Fisher-Rao metric).

 An even more fundamental physical meaning is assumed within the theory that we originally called Quantum Thermodynamics \cite{Frontiers,LectureNotes,HG,thesis,Cimento1,Cimento2} but more recently renamed Intrinsic Quantum Thermodynamics (IQT) to avoid confusion with the QT model just outlined.

IQT assumes that the second law of thermodynamics should complement the laws of mechanics even at the single particle level \cite{HG}. This can be done if we accept that the true individual quantum state of a system, even if fully isolated and uncorrelated from the rest of the
universe, requires density operators $\rho$ that are not
necessarily idempotent. Over the set of idempotent $\rho$'s, QT
coincides with Quantum Mechanics (QM), but it differs fundamentally
from QM because it assumes a broader set of possible states,
corresponding to the set of non-idempotent $\rho$'s. This way, the entropy
functional $S(\rho)$ becomes in IQT an intrinsic fundamental property.\footnote{In a sense it accomplishes the conceptual program, so intensely sought for also by Ilya Prigogine and coworkers \cite{Prigogine}, of answering the following questions \cite{Frontiers}: \textit{What if entropy, rather than a statistical, information theoretic, macroscopic or phenomenological concept, were an intrinsic property of matter in the same sense as energy is universally understood to be an intrinsic property of matter? What if irreversibility were an intrinsic feature of the fundamental dynamical laws obeyed by all physical objects, macroscopic and microscopic, complex and simple, large and small? What if the second law of thermodynamics, in the hierarchy of physical laws, were at the same level as the fundamental laws of mechanics, such as the great conservation principles?}
 When viewed from such extreme perspective, the IQT conceptual scheme remains today as \quot{adventurous} as it was acutely perceived by John Maddox in 1985 \cite{Nature}.}

 In the framework of IQT the SEA  equation of motion (\ref{5f}) for $\rho$ which results from the expression for $\Pi_\gamma$ we  construct in the next section represents a strong implementation of the  MEP principle at the fundamental quantum level and generalizes the original framework in which we developed the SEA formalism about 30 years ago by making it compatible, at least in the near-equilibrium limit with MNEQT.

 Even the brief discussion above shows clearly that the differences between QSM, QIT, QT, and IQT are important on the interpretational and conceptual levels. Nevertheless, it is also clear that they all share the same basic mathematical framework. Hence, we believe that the SEA dynamical model, which they share on the mathematical basis, can find in the different theories different physical interpretations and applications.

 \section*{STEEPEST-ENTROPY-ASCENT/MAXIMAL-ENTROPY-PRODUCTION DYNAMICS. UNIFIED VARIATIONAL FORMULATION FOR FRAMEWORKS A TO F}

 In the preceding sections we formulated the non-equilibrium problem in various different frameworks in a unifying way that allows us to represent their dissipative parts in a single formal way. In essence, the state is represented by an element $\gamma$ of a suitable vector space $ \mathcal{L} $ equipped with an inner product $ (\cdot|\cdot)$. The term in the dynamical equation for $\gamma$ which is responsible for  dissipative irreversible relaxation and hence entropy generation is another element $\Pi_\gamma$ of $ \mathcal{L} $ which determines the rate of entropy production according to the relation
 \be \Pi_S=(\Phi|\Pi_\gamma) \label{1g} \ee
 and the rates of production of the conserved properties $C_i$ according to the relation
  \be \Pi_{C_i}=(\Psi_i|\Pi_\gamma) \label{2g} \ee
  Except for the RET Framework D, where we have no explicit expressions for $\Phi$ and $\Psi_i$, in Frameworks A, B, C we found that $\Phi= -k(\ln \gamma^2)\, \gamma$ and $\Psi_i=2C_i \gamma$, in Framework F we found that $\Phi= -k(\ln \gamma\gamma^\dagger)\, \gamma$ and $\Psi_i=2C_i \gamma$.

 The formulation in terms of square roots of probabilities in Framework C, of the phase density in Frameworks A and B, of the density operator in Framework F takes care of the important condition that for the evolution law to be well defined it must conserve the nonnegativity of probabilities, phase densities and density operators (which must also remain self adjoint).

  Our next objective is to implement the MEP principle. We do this by assuming that the time evolution of the state $\gamma$ follows the path of steepest entropy ascent in $ \mathcal{L} $. So, for any given state $\gamma$, we must find the $\Pi_\gamma$ which maximizes the entropy production $\Pi_S$ subject to the constraints $\Pi_{C_i}=0$. But in order to identify the SEA path we are not interested in the unconditional increase in  $\Pi_S$ that we can trivially obtain by simply increasing the \quot{norm} of $\Pi_\gamma$ while keeping its direction fixed. Rather, the SEA path is identified by the direction of $\Pi_\gamma$ which maximizes $\Pi_S$ subject to the constraints, regardless of  norm of $\Pi_\gamma$. Hence, we must do the maximization at constant  norm of $\Pi_\gamma$.

  The norm of $\Pi_\gamma$ represents the square of the distance $d\ell$ traveled by $\gamma$ in the state space $ \mathcal{L} $ in the time interval $dt$, the square of the \quot{length} of the infinitesimal bit of path traveled in state space in the interval $dt$. The variational problem that identifies the SAE/MEP direction at each state $\gamma$ looks at all possible paths through $\gamma$, each characterized by a possible choice for $\Pi_\gamma$. Among all these paths it selects the one with the highest  entropy produced in the interval $dt$, $\Pi_S\,dt$ per unit of distance $d\ell$ traveled by $\gamma$.

  It is therefore apparent that we cannot identify a SAE/MEP path until we equip the space $ \mathcal{L} $ with a metric with respect to which to compute the distance $d\ell$ traveled and the norm of $\Pi_\gamma$.

  In our previous work \cite{ROMP},  we selected the Fisher-Rao metric based on the inner product $(\cdot|\cdot)$ defined on $ \mathcal{L} $. Indeed, in dealing with probability distributions it has been argued by several authors that the Fisher-Rao metric is the proper unique metric for the purpose of computing the distance between two probability distributions (see e.g. \cite{Wootters,Salamon,Braunstein}). According to this metric, the distance between two states $\gamma_1$ and $\gamma_2$ is given by \be d(\gamma_1,\gamma_2)=\sqrt{2}\arccos(\gamma_1|\gamma_2) \label{3g}\ee which implies that the distance traveled along a trajectory in state space is \be d\ell = 2\sqrt{(\Pi_\gamma|\Pi_\gamma)}\, dt \label{4g}\ee As a result, for Framework F the SEA dynamics we have originally proposed is most straightforward.

  However, here we will adopt a more general metric, which in Framework F generalizes our previous work and in the other frameworks provides a most general formulation. We assume the following expression for the distance traveled along a trajectory in state space \be d\ell = \sqrt{(\Pi_\gamma|\,\hat G\,|\Pi_\gamma)}\, dt \label{5g}\ee
  where $\hat G$ is a real, symmetric, and positive-definite operator on $ \mathcal{L} $ that we call the metric tensor, (super)matrix, or (super)operator depeding on the framework. In Framework F, since $ \mathcal{L} $ is the space of operators on the Hilbert space $\mathcal{H}$ of the quantum system, $\hat G$ is a superoperator on $\mathcal{H}$. However, a simple case is when $\hat G|A)=|GA)$ with $G$ some self-adjoint positive-definite operator in $ \mathcal{L} $.

   We may now finally state \textbf{the SAE/MEP variational problem} and solve it. The problem is to \textbf{find the instantaneous \quot{direction} of
   $\Pi_\gamma$ which maximizes the entropy production rate $\Pi_S$ subject to the constraints $\Pi_{C_i}=0$.} We solve it by maximizing the entropy production rate $\Pi_S$ subject to the constraints $\Pi_{C_i}=0$ and the additional constraint $(d\ell/dt)^2=\dot\epsilon^2=$ prescribed. The last constraint keeps the norm of $\Pi_\gamma$ constant so that we  maximize only with respect to its direction.  From Eq.\ (\ref{5g})  it amounts to keeping fixed the value of $(\Pi_\gamma|\,\hat G\,|\Pi_\gamma)$ at some small positive constant $\dot\epsilon^2$. The solution is easily obtained by the method of Lagrange multipliers. We seek the unconstrained maximum, with respect to $\Pi_\gamma$, of the Lagrangian
     \be  \Upsilon= \Pi_S - \sum_i \beta_i\, \Pi_{C_i} - \tau\, [(\Pi_\gamma|\,\hat G\,|\Pi_\gamma)-\dot\epsilon^2] \label{8g1} \ee
     where $\beta_i$ and $\tau$ are the Lagrange multipliers. They must be independent of $\Pi_\gamma$ but can be functions of the state $\gamma$.
Using Eqs.\ (\ref{1g}) and (\ref{2g}), we rewrite (\ref{8g1}) as follows
   \be  \Upsilon= (\Phi|\Pi_\gamma) - \sum_i \beta_i\, (\Psi_i|\Pi_\gamma) - \tau\, [ (\Pi_\gamma|\,\hat G\,|\Pi_\gamma)-\dot\epsilon^2] \label{8g} \ee
Taking the variational derivative of $\Upsilon$ with respect to $|\Pi_\gamma)$ and setting it equal to zero we obtain
  \be  \frac{\delta\Upsilon}{|\delta \Pi_\gamma)} = |\Phi) -\sum_i \beta_i\, |\Psi_i) -\tau \hat G|\Pi_\gamma) =0 \label{9g} \ee
  Thus, we obtain the SEA/MEP general evolution equation (the main result of this paper)
  \be |\Pi_\gamma)=\hat L\,|\Phi -\sum_j \beta_j\, \Psi_j)\label{10g} \ee
  where we define for convenience
  \be \hat L=\frac{1}{\tau}\hat G^{-1} \label{10g1}\ee
  Since in the various frameworks $\hat L$ can be connected with the generalized Onsager conductivity (super)matrix in the near equilibrium regime, we see here that  $\tau \hat L$ is the inverse of the metric (super)matrix $\hat G$ with respect to which the dynamics is SEA/MEP. In other words, denoting the generalized Onsager resistivity (super)matrix by $ \hat R$ we have: $ \hat R$ = $\tau\,\hat G$. Since, $\hat G$ is positive definite and symmetric, so are $\hat L$ and $ \hat R$. In other words, the SEA assumption entails Onsager reciprocity.

  Inserting Eq.\ (\ref{10g}) into the conservation constraints (\ref{2g}) yields the important system of equations which defines the values of the Lagrange multipliers $\beta_j$,
  \be  \sum_j  (\Psi_i|\,\hat L\,|\Psi_j)\,\beta_j =(\Psi_i|\,\hat L\,|\Phi)\label{11g} \ee
This system can be readily solved for the $\beta_j$'s (for example by Cramer's rule)  because the  functionals $(\Psi_i|\hat L|\Psi_j)$ and $(\Psi_i|\hat L|\Phi)$ are readily computable for the current state $ \gamma$. When Cramer's rule is worked out explicitly, the SEA equation (\ref{10g}) takes the form of a ratio of determinants with which we  presented it in the IQT framework \cite{thesis,Cimento1,Cimento2,ROMP}.

  We can now immediately prove the general consistence with the thermodynamic principle of entropy non-decrease ($H$-theorem in Framework B). Indeed, subtracting  Eqs.\ (\ref{2g}) each multiplied by the corresponding $\beta_j$'s  from Eq.\ (\ref{1g}) and then inserting Eq.\ (\ref{10g}) yields the following explicit expression for  rate of entropy production
  \bea \Pi_S&=&(\Phi|\Pi_\gamma) = (\Phi-\sum_j \beta_j\, \Psi_j |\Pi_\gamma) \nonumber\\
  &=&(\Phi-\sum_i \beta_i\, \Psi_i |\,\hat L\,|\Phi -\sum_j \beta_j\, \Psi_j)\ge 0 \label{12g}\eea
which is clearly nonnegative-definite by virtue, again, of the nonnegativity that must be assumed for a well defined metric superoperator $\hat G$.

It is interesting to write the expression for  the (prescribed) speed $d\ell/dt$ at which the state $\gamma$ evolves along the SEA/MEP path.  This amounts to inserting Eq.\ (\ref{10g}) into the additional constraint $(d\ell/dt)^2=\dot\epsilon^2=$ prescribed. We readily find
\bea \frac{d\ell^2}{dt^2}&=&(\Pi_\gamma|\,\hat G\,|\Pi_\gamma) \nonumber\\
  &=&\frac{1}{\tau^2}(\Phi-\sum_i \beta_i\, \Psi_i |\,\hat G^{-1}\hat G\hat G^{-1}\,|\Phi -\sum_j \beta_j\, \Psi_j)\label{13g1}\\ &=&\frac{1}{\tau}\Pi_S = \dot\epsilon^2\label{13g} \eea
 from which we see that the Lagrange multiplier $\tau$ is related to the entropy production rate and the speed $d\ell/dt$. In other words, through $\tau$ we may specify either the speed at which $\gamma$ evolves along the SEA/MEP trajectory in state space or the instantaneous rate of entropy production. Indeed, using Eq.\ (\ref{13g1}), we obtain
 \bea \tau&=&\frac{\sqrt{(\Phi-\sum_i \beta_i\, \Psi_i |\,\hat G^{-1}\,|\Phi -\sum_j \beta_j\, \Psi_j)}}{d\ell/dt}\label{14g1}\\ &=&\frac{(\Phi-\sum_i \beta_i\, \Psi_i |\,\hat G^{-1}\,|\Phi -\sum_j \beta_j\, \Psi_j)}{\Pi_S}\label{14g} \eea
 Hence, using $\tau$ given by Eq.\ (\ref{14g}) the evolution equation  Eq.\ (\ref{10g}) will produce a SEA/MEP trajectory in state space with the prescribed entropy production $\Pi_S$. Eq.\ (\ref{14g}) also clearly supports the interpretation of $\tau$ as the \quot{overall relaxation time}.

In general, we may  interpret the vector
\be |\Lambda)= \hat G^{-1/2}\,|\Phi-\sum_i \beta_i\, \Psi_i)\label{18g}\ee
as a vector of \quot{generalized partial affinities}. In terms of this vector, Eq.\ (\ref{10g}) rewrites as
 \be\hat G^{1/2}\, |\Pi_\gamma)=\frac{1}{\tau}\,|\Lambda)\label{19} \ee
When only some of the partial affinities in the vector $\Lambda$ are zero, the state is partially equilibrated (equilibrated with respect to the corresponding underlying components of the state $\gamma$). When the entries of the vector $\Lambda$  are all zero, then and only then we have an equilibrium state or a non-dissipative limit cycle. In fact, that is when and only when the entropy production vanishes. $(\Lambda|\Lambda)$, which with respect to the metric tensor $\hat G$ is the norm of the vector  $\Phi -\sum_j \beta_j\, \Psi_j$, represents a measure of the \quot{overall degree of disequilibrium}. It is important to note that this definition is valid no matter how far the state is from the (maximum entropy) stable equilibrium state, i.e., also for highly non-equilibrium states.

 Eq.\ (\ref{14g}) rewrites as
\be \Pi_S = \frac{(\Lambda|\Lambda)}{\tau}\label{20g}\ee
which shows that the rate of entropy production is proportional to the overall degree of disequilibrium. The relaxation time $\tau$ may be a state functional and needs not be constant, but even if it is, the SEA principle provides a  nontrivial non-linear evolution equation that is well defined and reasonable even far from equilibrium.

 We finally note that when the only contribution to the entropy change comes from the production term $\Pi_S$ (for example in Framework B in the case of homogeneous relaxation in the absence of entropy fluxes, or in Framework F for an isolated system), i.e., when the entropy balance equation  reduces to $dS/dt=\Pi_S$, Eq.\ (\ref{13g} ) may be rewritten as
  \be \frac{d\ell}{dt/\tau}=\frac{dS}{d\ell}\label{15g} \ee
  from which we see that when time $t$ is measured in units of $\tau$ the "speed" along the SEA trajectory  is equal to the local entropy gradient  along the trajectory.

  If the state $\gamma$ moves only due to the dissipative term $\Pi_\gamma$ (for example in Framework F when $[H,\gamma\gamma^\dagger]=0$), then the overall length of the trajectory in state space traveled between $t=0$ and $t$ is  given by
  \be \ell(t) = \int_0^t \sqrt{(\Pi_\gamma|\,\hat G\,|\Pi_\gamma)}\, dt \label{16g}\ee
  and, correspondingly, we may also define the \quot{non-equilibrium action}
   \be \Sigma = \frac{1}{2}\int_0^t (\Pi_\gamma|\,\hat G\,|\Pi_\gamma)\, dt =\frac{1}{2}\int_0^t \frac{\Pi_S}{\tau}\, dt =\frac{1}{2}\int_0^t \frac{(\Lambda|\Lambda)}{\tau^2}\, dt\label{17g}\ee
   where for the last two equalities we used Eq.\ (\ref{13g}) and Eq.\ (\ref{20g}), respectively.

The explicit expressions of the SEA/MEP dynamical equations that result in the six different frameworks treated here can be readily obtained but will be given   elsewhere.

\section*{PICTORIAL REPRESENTATIONS}

Let us  give pictorial representations of the vectors that we defined in the SEA/MEP construction. We consider first the simplest scenario of a uniform metric tensor $\hat G = \hat I$.

\begin{figure}[t]
      \centering
       \includegraphics[width=0.5\textwidth]{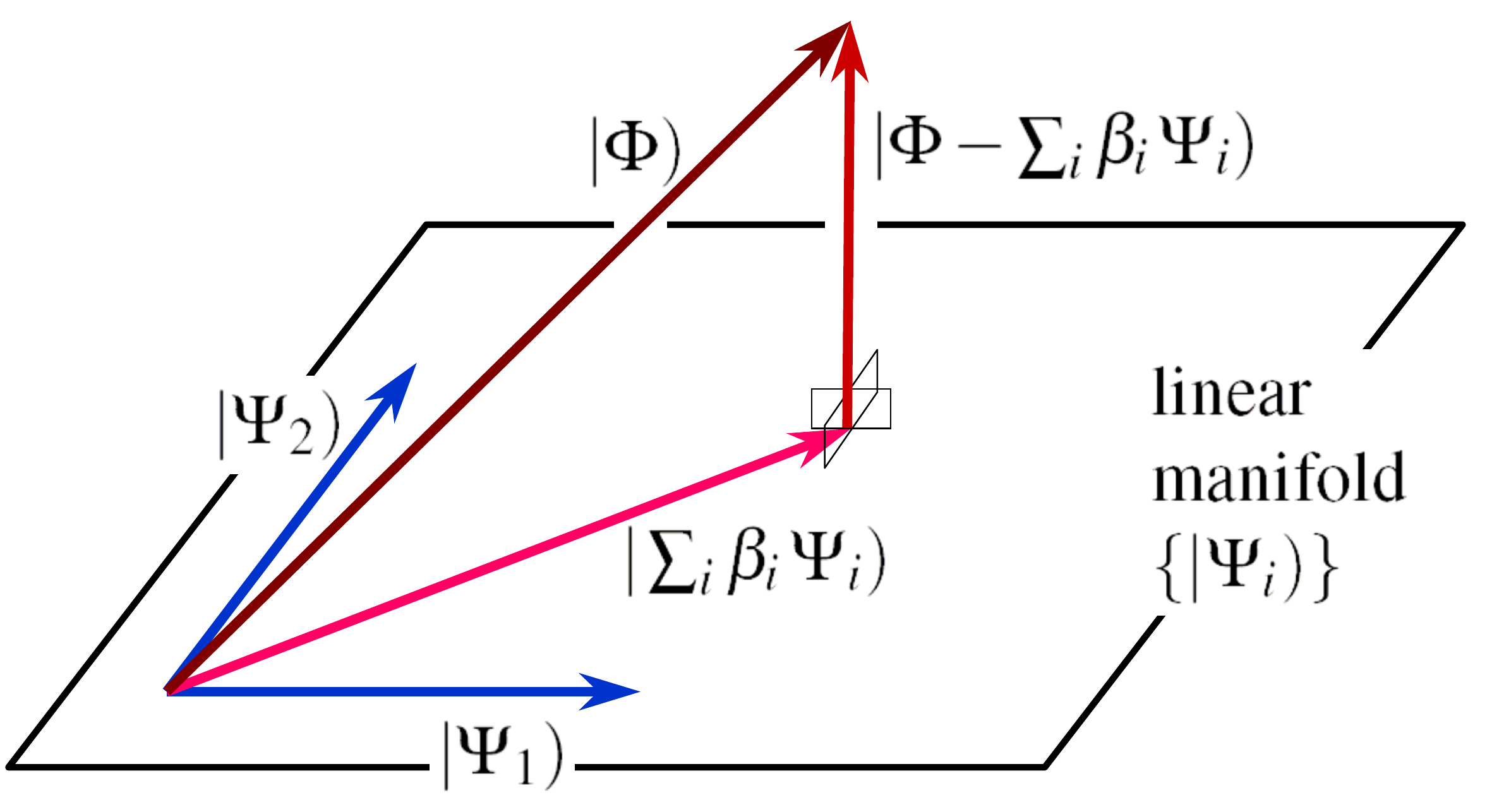}
       \caption{\label{Figure4}Pictorial representation of the linear manifold spanned by the vectors $|\Psi_i)$ and the orthogonal projection of $|\Phi)$ onto this manifold which defines the Lagrange multipliers $\beta_i$ in the case of a uniform metric $\hat G = \hat I$. The construction defines also the generalized affinity vector, which in this case is $|\Lambda)=| \Phi-\sum_i \beta_i\, \Psi_i)$.}
   \end{figure}

 \begin{figure}[t]
      \centering
       \includegraphics[width=0.5\textwidth]{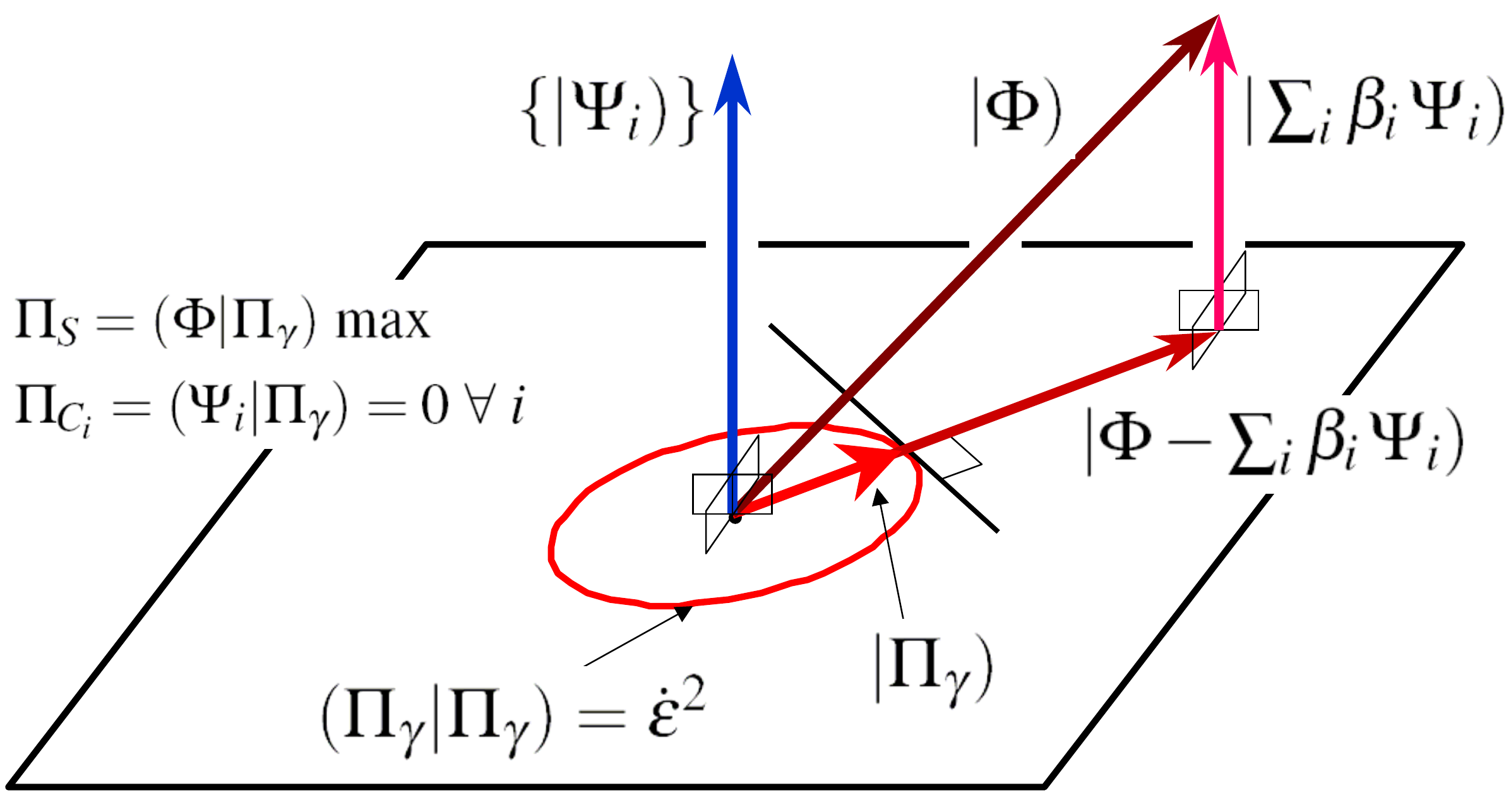}
       \caption{\label{Figure1}Pictorial representation of the SEA/MEP variational construction in the case of a uniform metric $\hat G = \hat I$. The circle represents the condition $(\Pi_\gamma|\Pi_\gamma)=\dot\epsilon^2$. The vector $|\Pi_\gamma)$ must be orthogonal to the $|\Psi_i)$'s in order to satisfy the conservation constraints $\Pi_{C_i}=(\Psi_i|\Pi_\gamma)=0$. In order to  maximize the scalar product  $(\Phi-\sum_i \beta_i\, \Psi_i|\Pi_\gamma)$, $|\Pi_\gamma)$ must have the same direction as $| \Phi-\sum_i \beta_i\, \Psi_i)$.}
   \end{figure}

Figure \ref{Figure4} gives a pictorial representation of the linear manifold spanned by the vectors $|\Psi_i)$'s and the orthogonal projection of $|\Phi)$ which defines the Lagrange multipliers $\beta_i$ in the case of uniform metric, i.e., the orthogonality conditions  $(\Psi_j|\Phi-\sum_i \beta_i\, \Psi_i)=0$ for every $j$, which is Eq.\ (\ref{11g}) with $\hat L = \hat I$. The construction defines also the generalized affinity vector, which in this case is $|\Lambda)=| \Phi-\sum_i \beta_i\, \Psi_i)$ and is orthogonal to the linear manifold spanned by the vectors $|\Psi_i)$'s.

Figure \ref{Figure1} gives a pictorial representation of the subspace orthogonal to the linear manifold spanned by the $|\Psi_i)$'s that here we denote for simplicity by $\{ |\Psi_i) \}$. The vector $|\Phi)$ is decomposed into its component $|\sum_i \beta_i\, \Psi_i)$ which lies in $\{ \Psi_i \}$  and its component $|\Phi-\sum_i \beta_i\, \Psi_i)$ which lies in the orthogonal  subspace.

The circle in Figure \ref{Figure1} represents the condition $(\Pi_\gamma|\Pi_\gamma)=\dot\epsilon^2$ corresponding in the uniform metric to the  prescribed rate of advancement in state space, $\dot\epsilon^2=(d\ell/dt)^2$. The compatibility with the conservation constraints $\Pi_{C_i}=(\Psi_i|\Pi_\gamma)=0$ requires that  $|\Pi_\gamma)$ lies in subspace orthogonal to the  $|\Psi_i)$'s. To take the SEA
the direction $|\Pi_\gamma)$ must maximize the scalar product  $(\Phi-\sum_i \beta_i\, \Psi_i|\Pi_\gamma)$. This clearly happens when $|\Pi_\gamma)$ has the same direction as the vector $| \Phi-\sum_i \beta_i\, \Psi_i)$ which in the uniform metric coincides with  the generalized affinity vector $|\Lambda)$.

   \begin{figure}[t]
      \centering
       \includegraphics[width=0.5\textwidth]{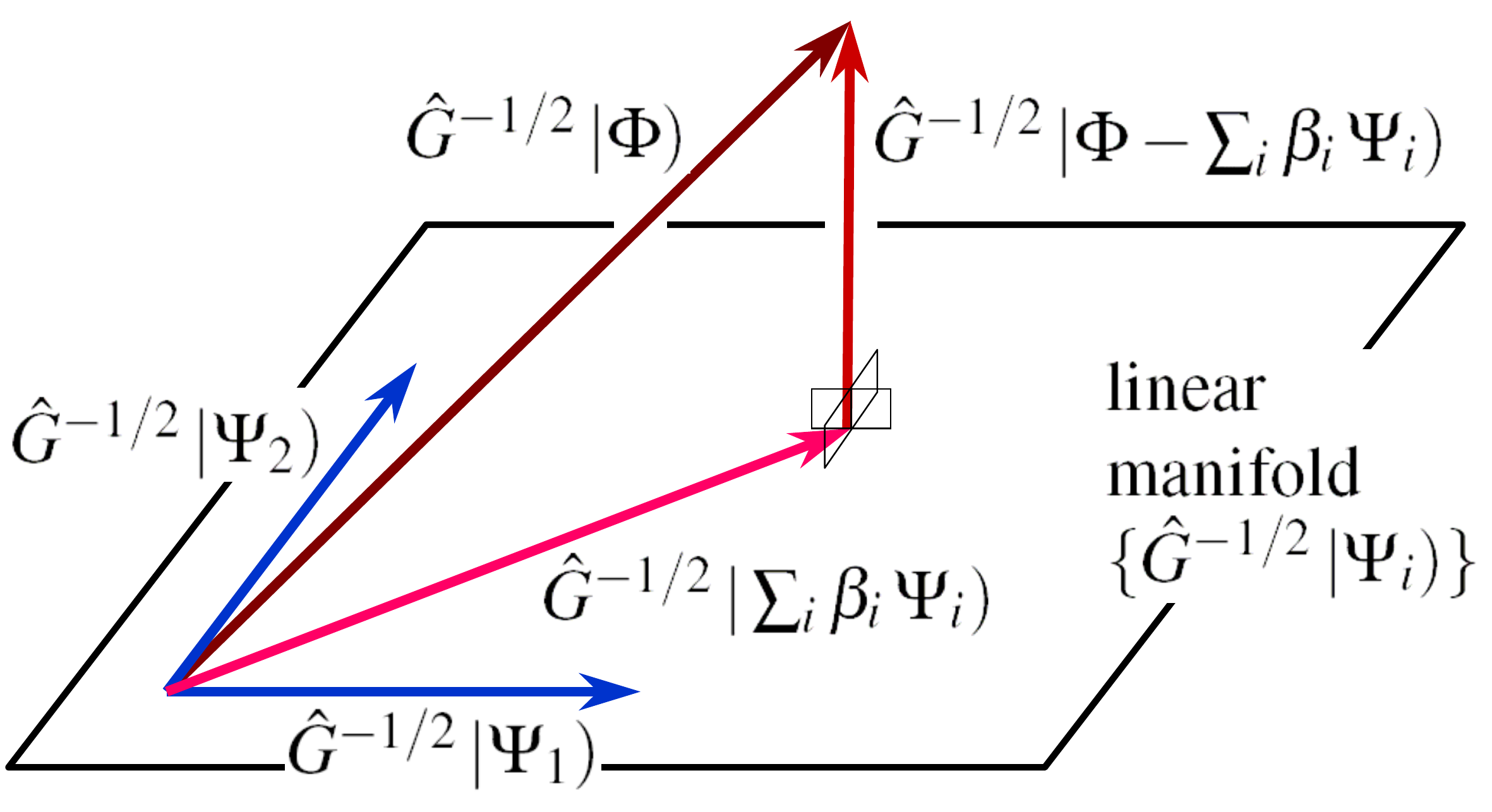}
       \caption{\label{Figure5}Pictorial representation of the linear manifold spanned by the vectors $\hat G^{-1/2}\,\Psi_i$ and the orthogonal projection of $\hat G^{-1/2}\,|\Phi)$ onto this manifold which defines the Lagrange multipliers $\beta_i$ in the case of  a non-uniform metric $\hat G$. The construction defines also the generalized affinity vector $|\Lambda)=\hat G^{-1/2}\,| \Phi-\sum_i \beta_i\, \Psi_i)$.}
   \end{figure}
 \begin{figure}[t]
      \centering
       \includegraphics[width=0.5\textwidth]{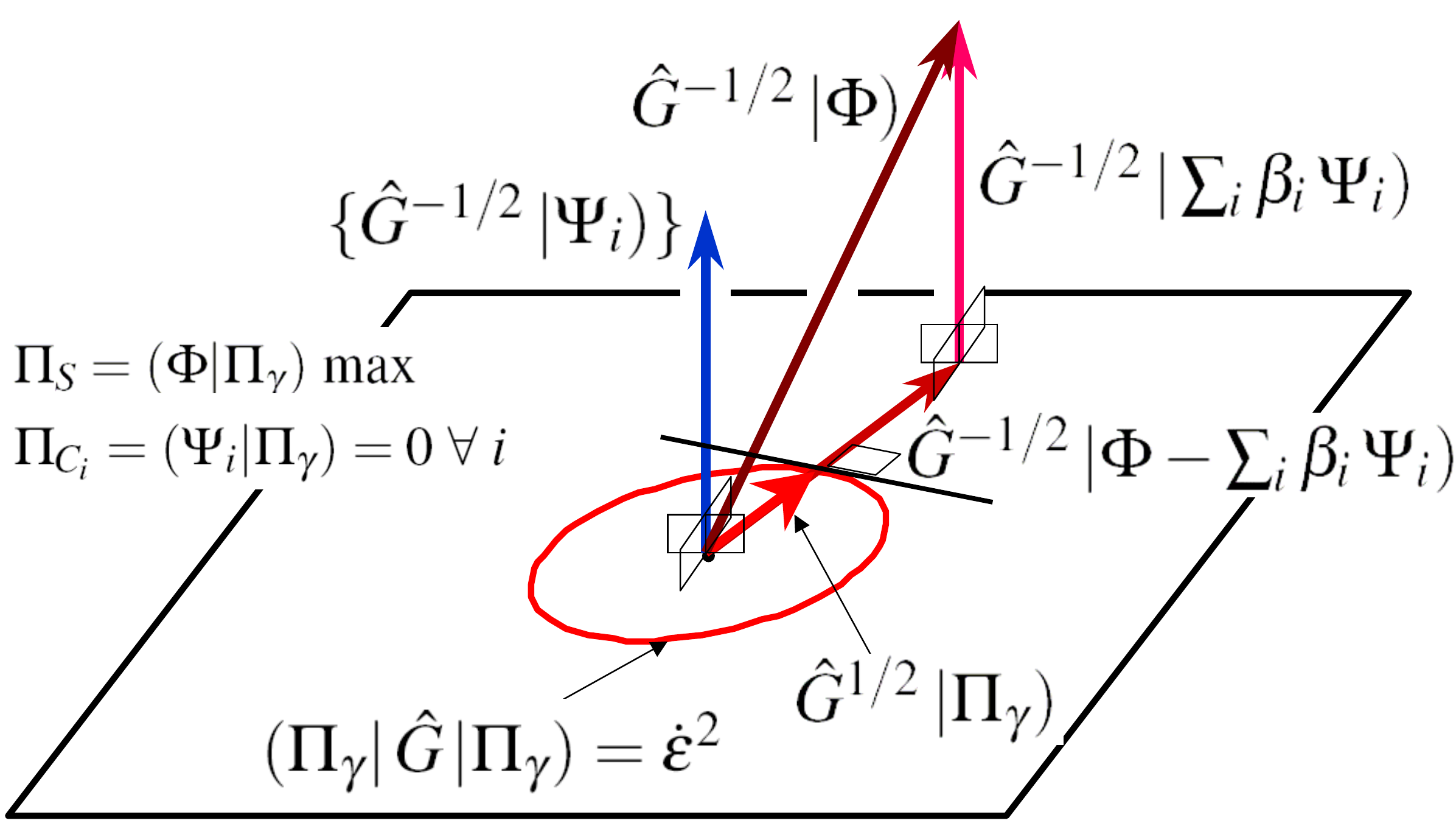}
       \caption{\label{Figure3}Pictorial representation of the SEA/MEP variational construction in the case of a non-uniform metric $\hat G$. The circle  represents  the  condition $(\Pi_\gamma|\,\hat G\,|\Pi_\gamma)=\dot\epsilon^2$, corresponding to the norm of vector $\hat G^{1/2}\,|\Pi_\gamma)$. This vector must be orthogonal to the $\hat G^{-1/2}\,|\Psi_i)$'s in order to satisfy the conservation constraints $\Pi_{C_i}=(\Psi_i|\Pi_\gamma)=0$. In order to  maximize the scalar product  $\Pi_{S}=(\Phi|\Pi_\gamma)=(\Phi-\sum_i \beta_i\, \Psi_i|\Pi_\gamma)$, vector $\hat G^{1/2}\,|\Pi_\gamma)$ must have the same direction as $|\Lambda)=\hat G^{-1/2}\,| \Phi-\sum_i \beta_i\, \Psi_i)$.}
   \end{figure}

Next, we consider the more general  scenario of a non-uniform metric tensor $\hat G$.
Figure \ref{Figure5} gives a pictorial representation of the linear manifold spanned by the vectors $\hat G^{-1/2}\,|\Psi_i)$ and the orthogonal projection of $\hat G^{-1/2}\,|\Phi)$ which defines the Lagrange multipliers $\beta_i$ in the case of non-uniform metric $\hat G$, where the orthogonality conditions that define the $\beta_i$'s are $(\Psi_j|\,\hat G^{-1}\,|\Phi-\sum_i \beta_i\, \Psi_i)=0$ for every $j$, which is Eq.\ (\ref{11g}).  The construction defines also the generalized affinity vector $|\Lambda)=\hat G^{-1/2}\,| \Phi-\sum_i \beta_i\, \Psi_i)$ which  is orthogonal to the linear manifold spanned by the vectors $\hat G^{-1/2}\,|\Psi_i)$'s.

Figure \ref{Figure3} gives a pictorial representation of the subspace orthogonal to the linear manifold spanned by the $\hat G^{-1/2}\,|\Psi_i)$'s that here we denote for simplicity by $\{ \hat G^{-1/2}\,|\Psi_i) \}$. The vector $\hat G^{-1/2}\,|\Phi)$ is decomposed into its component $\hat G^{-1/2}\,|\sum_i \beta_i\, \Psi_i)$ which lies in $\{ \hat G^{-1/2}\,|\Psi_i ) \}$  and its component $|\Lambda)=\hat G^{-1/2}\,|\Phi-\sum_i \beta_i\, \Psi_i)$ which lies in the orthogonal subspace.

The circle in Figure \ref{Figure3} represents  the more general condition $(\Pi_\gamma|\,\hat G\,|\Pi_\gamma)=\dot\epsilon^2$ corresponding in the non-uniform metric to the  prescribed rate of advancement in state space, $\dot\epsilon^2=(d\ell/dt)^2$. It is clear that the  direction of $\hat G^{1/2}\,|\Pi_\gamma)$ which maximizes the scalar product  $(\Phi-\sum_i \beta_i\, \Psi_i|\Pi_\gamma)$, is when $|\Pi_\gamma)$ is in the direction of the point of tangency between the ellipse and a line orthogonal to $| \Phi-\sum_i \beta_i\, \Psi_i)$.

The compatibility with the conservation constraints $\Pi_{C_i}=(\Psi_i|\Pi_\gamma)=0$ requires that  $\hat G^{1/2}\,|\Pi_\gamma)$ lies in subspace orthogonal to the  $\hat G^{-1/2}\,|\Psi_i)$'s. To take the SEA/MEP
 direction, the vector $\hat G^{1/2}\,|\Pi_\gamma)$ must maximize the scalar product  $(\Phi-\sum_i \beta_i\, \Psi_i|\Pi_\gamma)$, which is equal to the entropy production $\Pi_{S}=(\Phi|\Pi_\gamma)$ since $(\Psi_i|\Pi_\gamma)=0$. This clearly happens when $|\Pi_\gamma)$ has the same direction as the generalized affinity vector $|\Lambda)=\hat G^{-1/2}\,| \Phi-\sum_i \beta_i\, \Psi_i)$.

\section*{CONCLUSIONS}

 In this paper, we review the essential mathematical elements of the formulations of six different approaches to the description of non-equilibrium dynamics. At the price of casting some of them in a somewhat unusual notation, we gain the possibility to set up a unified formulation, which allows us to investigate the locally Maximum Entropy Production (MEP) principle in all these contexts. It is a generalization to non-homogeneous cases of the local Steepest Entropy Ascent (SEA) concept whereby the time evolution the   state  is assumed to follows a path in state space which, with respect to an underlying metric, is always tangent to the direction of maximal entropy production compatible with the conservation constraints.
%

   The present SEA/MEP unified formulation allows us to extend at once to all these frameworks  the SEA concept which has so far been considered only in the framework of quantum thermodynamics. Actually, the present formulation constitutes a generalization even in the quantum thermodynamics framework and constitutes a natural generalization to the far-nonequilibrium domain of Mesoscopic Non-Equilibrium Quantum Thermodynamics.

   The  analysis emphasizes that in the SEA/MEP implementation of the MEP principle, a key role is played by the geometrical metric with respect to which to measure the length of a trajectory in state space. The metric tensor turns out to be directly related to the inverse of the Onsager's generalized conductivity tensor.

   We  conclude that in most of the existing theories of non-equilibrium the time evolution of the state representative can be seen to actually follow in state space the path of SEA with respect to a suitable metric connected with the generalized conductivities. This is true in the near-equilibrium limit, where in all frameworks it is possible to show that the traditional assumption of linear relaxation coincides with the SEA/MEP result.  Since the generalized conductivities represent, at least in the near-equilibrium regime, the strength of the system's reaction when pulled out of equilibrium, it appear that their inverse, i.e., the generalized resistivity tensor, represents the metric with respect to which the time evolution, at least in the near equilibrium, is SEA/MEP.

   Far from equilibrium  the resulting unified family of  SAE/MEP dynamical models is a very fundamental as well as practical starting point because it features an intrinsic  consistency with the second law of thermodynamics. The proof of nonnegativity of the local entropy production density is a general and straightforward regardless of the details of the underlying metric tensor. In a variety of fields of application, the present unifying approach may prove useful in providing a new basis for effective numerical and theoretical models of irreversible, conservative relaxation towards equilibrium from far non-equilibrium states.

\section*{ACKNOWLEDGMENTS}

\noindent The author gratefully acknowledges the
Cariplo--UniBS--MIT-MechE faculty exchange program co-sponsored by
UniBS and the CARIPLO Foundation, Italy under grant 2008-2290. This work is part of EOARD (European Office of Aerospace R\&D)  grant FA8655-11-1-3068 and italian MIUR  PRIN-2009-3JPM5Z-002.

\bibliographystyle{unsrt}

    \end{document}